# Highly Efficient, Bright and Stable Colloidal Quantum Dot Short-Wave Infrared Light Emitting Diodes

Santanu Pradhan[1], Mariona Dalmases[1], Ayse-Bilgehan Baspinar[1] and Gerasimos Konstantatos[1, 2]*

[1]ICFO-Institut de Ciencies Fotoniques, The Barcelona Institute of Science and Technology, 08860 Castelldefels (Barcelona), Spain

[2]ICREA—Institució Catalana de Recerca i Estudis Avançats, Passeig Lluís Companys 23, 08010 Barcelona, Spain

* gerasimos.konstantatos@icfo.es

*Abstract:* Unbalanced charge injection is deleterious for the performance of colloidal quantum dot (CQD) light emitting diodes (LEDs) as it deteriorates the quantum efficiency (QE), brightness and operational lifetime. CQD LEDs emitting in the infrared have previously achieved high quantum efficiencies but only when driven to emit in the low radiance regime. At higher radiance levels, required for practical applications, the efficiency decreased dramatically in view of the notorious efficiency droop. Here we report a novel methodology to regulate charge supply in multinary bandgap CQD composites that facilitates improved charge balance. Our approach is based on engineering the energetic potential landscape at the supra-nanocrystalline level that has allowed us to report short-wave infrared (SWIR) PbS CQD LEDs with record-high external QE in excess of 8%, most importantly, at a radiance level of ~ 5 $Wsr^{-1}m^2$, an order of magnitude higher than prior reports. Furthermore, the balanced charge injection and Auger recombination reduction has led to unprecedentedly high operational stability with radiance half-life of 26068 hours at a radiance of $1 Wsr^{-1}m^{-2}$.

# 1. Introduction

Light emission in the short wave infrared is of paramount importance for a large spectrum of applications including environmental sensing, surveillance, night and automotive vision, food and product quality control, on-chip spectroscopy, biomedical imaging and 3D imaging [1-4]. There is therefore an emerging need for low cost, CMOS compatible SWIR LED solutions with performance competitive to that of currently existing technologies based on rigid, non-CMOS and costly III-V semiconductors. Colloidal quantum dots (CQDs), in particular those based on lead-chalcogenides, stand out as the most promising candidates currently to fulfill the required specifications [5]. In view of this opportunity, there has been lately recent progress on infrared lead sulphide (PbS) CQD LEDs with notable increase in their external quantum efficiency, an important figure of merit for LEDs, from ~4% employing core-shell CQDs [6] or CQD-in-perovksite structures [7] to a recent record of 7.8% based on multinary bandgap CQD active layers [8]. Such values of EQE are now comparable to commercial state-of-the-art InGaAs LEDs [9]. Yet for practical applications, another equally important figure of merit is the level of brightness that such EQE values can be achieved at. The LED should emit light with high quantum efficiency and high brightness. Previously, such EQE values for infrared PbS CQD LEDs have been reported at low radiance levels, i.e. in the low injection current regime. The reason is the notorious charging effects that take place in LEDs upon unbalanced charge injection which in turn deteriorate drastically the quantum efficiency, an effect that has been well-known as the LED droop. The origin of such droop in efficiency is due to Auger recombination [10] or formation of charged exciton species [11, 12] that outcompetes radiative recombination or the formation of space charge regions that further inhibit balanced injection of carriers [13]. To address this problem several approaches have been taken for CQD LEDs emitting in the visible aiming to regulate charge injection by the use of blocking layers [14] or reducing Auger recombination with core-shell structures [15] and graded core-shell structures [12]. However, infrared PbS CQD LEDs still suffer from the large droop effect putting an important barrier towards their commercialization. Here we report on a new methodology to suppress droop in infrared PbS CQD LEDs by engineering the energetic potential landscape of the active layer at the supra-nanocrystalline level. We employ binary and ternary blends of CQDs in which we judiciously control the band offsets between the matrix CQDs and the emitter CQDs (Fig. 1a-c) in order to regulate and balance the supply of electrons and holes in the emitter CQDs. By doing so we achieve record EQE performance at high radiance levels that had not been reached before by CQD LEDs emitting in the short-wave infrared.

## 2. Results and Discussion

The blended CQD based binary devices comprise two types of PbS QDs with different bandgaps [8, 16]. The larger bandgap ($E_{gM}$) QDs serve as the carrier supply matrix and the smaller-bandgap ($E_{gE}$) QDs act as the emitters. The injected charges transverse through the matrix and recombine radiatively at the emitter QD sites. The schematic band diagram of the device is shown in Fig. 1a (detail of the device preparation is mentioned in the experimental section). The optimized performance was obtained with a 7.5% mixing of emitter QDs in the blend [8]. The band levels of the matrix QDs determine at large the supply of electrons and holes in the emitter QDs as shown in Fig. 1b & 1c. The barrier heights for electron and hole injection are mainly determined by the conduction and valence band offset with the respective electrodes. To test this hypothesis, we fabricated and characterized binary-blend devices varying the bandgap of the matrix QDs (by varying the QD size), without altering the emitter QD bandgap in the blend. We have selected matrix QDs with three different sizes corresponding to bandgap values of 1.75, 1.5 and 1.32 eV whereas the bandgap of the emitter QDs has been kept constant at 0.95 eV. For clarity, we consider the value of the bandgap here as the excitonic absorption peak of the QDs in solution. The electroluminescence spectra of the devices are shown in Fig. S1. The spectral shape and emission peak position do not change with the increasing applied current indicating the high device stability. Figure 1d shows the voltage bias dependent radiance plot for devices with varying the matrix QD bandgap. The LED employing the 1.75 eV matrix QD based device shows larger turn-on voltage (~0.7 V) compared to the other two devices (0.55 V- 0.6 V) and overall lower radiance in the low injection regime. This is suggestive of a carrier injection barrier presence between the electrodes and the high bandgap matrix QDs. However the situation is reversed in the high injection regime: there the LED with the 1.75 eV matrix QD demonstrates increasing radiance upon bias reaching a radiance level of ~35 $Wsr^{-1}m^{-2}$ at high applied bias of 5V, whereas the LEDs based on smaller $E_{gM}$ QDs demonstrate saturation of emission, a strong indication of droop onset. This becomes evident in Fig. 1e where we plot the EQE as a function of injected current density (the corresponding plot as a function of applied bias is shown in Fig. S2a). First, we note a gradual improvement of the overall EQE with increasing bandgap of the QD matrix. The peak EQE increases from 1.8% for 1.32 eV $E_{gM}$ QD to 4.3% for 1.5 eV $E_{gM}$ QD based device and further to 6.6% for 1.75 eV based device. Most importantly, we note a gradual improvement of EQE droop with the use of higher bandgap QDs. The $J_{1/2EQE}$ (defined as the injected current density for which the EQE becomes half of the peak value) increases form 76 $mAcm^{-2}$ for 1.5 eV $E_{gM}$ device to 330 $mAcm^{-2}$ for 1.75 eV $E_{gM}$ QD based device. Figure 1f summarizes the achievement of this work by plotting the EQE of the devices as a function of their radiance. LEDs with 1.32 eV $E_{gM}$ and 1.5 eV $E_{gM}$ QD based device show much higher EQE droop in the high radiance regime compared to the 1.75 eV $E_{gM}$ QD based device. In fact, 1.75 eV $E_{gM}$ QD based device exhibit nearly constant EQE (within the 90% of peak EQE) from 0.3 $Wsr^{-1}m^{-2}$ to 5 $Wsr^{-1}m^{-2}$ while the other devices show a drastic drop of performance upon increasing radiance (comparisons of different matrix QDs and emitter QDs based device performance are shown in Fig S3 & S4). The improvement of EQE in the higher current injection regime also results in the increment of PCE (power conversion efficiency) defined as the ratio of applied electrical power to the emitted optical power of the devices (Fig. S2b). To our knowledge, these devices demonstrate the highest EQE along with the lowest EQE droop for CQD LEDs emitting in the short wave infrared (Table S4).

To gain further insights on the role of the matrix bandgap in response to our experimental observations, namely the improvement in the EQE and reduction in EQE droop for higher bandgap matrix QD devices, we have performed some SCAPS simulations. In SCAPS, we have taken the blend devices as the series of heterojunctions between the matrix and the emitter QDs keeping the emitter to matrix ratio intact as the experimental value (supporting information S5). The EQE of a LED device can be express as [7, 17],

$$EQE = \eta_{PL}\chi\eta_{tr}\eta_{em} \qquad (1)$$

Where $\eta_{PL}$ is the Photoluminescence quantum yield (PLQY) of the active material. $\chi$ is the fraction of spin allowed transition (for QDs, it is taken as 1). $\eta_{tr}$ is the charge transport efficiency which is a combined effect of charge injection efficiency ($\eta_{inj}$) and charge diffusion in the active layer ($\eta_{diff}$). $\eta_{em}$ is the emission efficiency, depends on the optical properties of the device substrate. We have simulated the variation of $\eta_{PL}$ and $\eta_{tr}$ as a function of band-offset between the matrix and the emitter QDs (keeping the emitter QD bandgap fixed and varying the matrix bandgap) (Fig. 2a). It is to be noted that $\eta_{em}$ does not depend on the active layer variation and thus we assume it plays insignificant role in EQE variation in our case. We have simulated the $\eta_{PL}$ by calculating the radiative recombination ($R_{rad}$) and total recombination ($R_{total}$) (combination of radiative, SRH (trap assisted) and Auger recombination). Figure 2b & 2c showed the variation of $R_{rad}$ and $R_{total}$ as a function of conduction and valence band offset. The simulation showed that for a particular applied voltage, the $R_{rad}$ and $R_{total}$ are mainly influenced by the valence band off-set. The influence of conduction band off-set on $R_{rad}$ and $R_{total}$ is relatively small. This is likely due to the mismatch of excess electrons and lower hole quantity in the emitter sites. The change of $\Delta E_C$ only changes the quantity of injected electrons - that is already excess in number compared to the number of holes so there will not be a significant change in $R_{rad}$, although the excess of electrons has a significant influence on the Auger recombination (Fig. S6). Then we have compared the values for two matrix QDs of interest (1.75 and 1.5 eV). The band values were taken from ultraviolet photoelectron spectroscopy (UPS) measurements (Supporting information S7). Simulation shows nearly similar value of $R_{rad}$ and $R_{total}$ for 1.75 and 1.5 eV $E_{gM}$ QD based devices (considering the fact that the influence of Auger recombination in both the devices is not highly significant to change $R_{total}$). We have also observed a nearly similar PLQY for both 1.75 eV (54±5)% and 1.5 eV (51±5)% based devices that supports the trend showed in the simulation (supporting information S8). Further, we have simulated the injection current as a function of band offset (Fig. 2d). It shows that the injection current decreases with increase of $\Delta E_C$ and $\Delta E_V$ indicating the effect of energy barrier on charge injection from the respective electrodes. We also have observed this effect experimentally as increasing the matrix QD band gap regulates the charge injection in the active layer (Fig. S2c). The injection efficiency (defined as the ratio of $R_{total}$ and charge injection rate ($I_{inj}/q$)) [17] showed a significant improvement for 1.75 eV based device compared to 1.5 eV $E_{gM}$ QD based device (Fig. 2e). Thus, the improvement in device performance is mainly coming from better charge balance and higher charge injection efficiency. Moreover, we have performed the simulation at a higher charge injection regime (applied bias 5V). The results are summarized in Fig. S7. This shows the difference of injection efficiency even higher at a higher applied bias for higher bandgap matrix based device. This is partly due to the improvement of $R_{rad}$ and partly due to the decrease of injection charges. This supports our experimental trend of achieving higher EQE and lower efficiency droop for higher bandgap matrix based device in the high injection regime. To corroborate our statements about the improved charge balance, we have estimated the hole and electron mobilites of the binary blends based on 1.5 and 1.75 eV $E_{gM}$ using space charge limited current technique (supporting information S9). 1.75 eV QD based blend showed better mobility balance between electrons and holes compared to the 1.5 eV QD based blend confirming the better charge balance and lower probability of Auger recombination with the higher bandgap matrix QDs [13].

Having reached compelling performance in terms of EQE and efficiency droop in binary blend devices upon engineering the bandgap of the matrix, next we wanted to test our method with ternary blend devices that have previously delivered record EQE [8]. The ternary blend further comprises some amount of ZnO NCs that serves to electronically passivate defects in the PbS CQDs [8] leading therefore to very high PLQY in solid state films and record EQE value of 7.8%. Figure 3a shows the schematic of the ternary blend where ZnO NCs act as electronic passivant for the PbS CQDs. Figure 3b compares the EQE of the ternary blend device with the binary blend device employing the optimized bandgap matrix QDs of 1.75 eV. In agreement with prior observations, the ternary blend device outperforms the binary counterpart and has currently reached record EQE value of 8.1%. The improved droop is further demonstrated in the ternary blend device by employing the optimized bandgap of the QD matrix as shown in Fig. 3c. The optimized ternary device delivers EQE of 8% at a higher radiance level range of 2-10 $Wm^{-2}sr^{-1}$ compared to the binary devices with similar $E_{GM}$ (peak EQE of 6.6% around 0.5-2 $Wm^{-2}sr^{-1}$). Furthermore, while comparing the EQE dependence on the radiance of ternary devices based on 1.5 and 1.75 eV $E_{gM}$, higher $E_{GM}$ QD based device showed better EQE in high radiance region (Fig. 3d). While ternary devices based on 1.75 eV $E_{gM}$ showed EQE ~8% at a radiance range of 2-10 $Wm^{-2}sr^{-1}$, 1.5 eV $E_{gM}$ QD based ternary devices[8] showed peak EQE ~7.7% at an order of magnitude lower radiance of 0.2-0.7 $Wm^{-2}sr^{-1}$.

Not only does unbalanced carrier injection in LEDs deteriorate performance, it also deteriorates the device lifetime. Charging in CQD LEDs has been identified as the main origin of device degradation [10] so its suppression is expected to improve also the stability of the devices.

Having improved charge balance in our devices we wanted to test this hypothesis also on their stability. Figure 4 shows the radiance as a function of time with a constant applied current. The devices were measured in ambient air conditions and without any encapsulation in place. The optimized 1.75 eV $E_{gM}$ QD based LED shows a much better stability compared to the one based on 1.5 eV $E_{gM}$. For a 7 $Wsr^{-1}m^{-2}$ initial radiance, the $t_{1/2}$ (defined as the time taken to reach the half of the initial radiance) increases from 42 h for 1.5 eV $E_{gM}$ QD based device to 532 h for 1.75 eV $E_{gM}$ QD based devices in ambient conditions (60% humidity and 298K temperature). Empirical calculations following a previous report [12] (supporting information S10) shows the expected $t_{1/2}$ for 1.75 eV $E_{gM}$ QD based devices increases to 26,068 h when the initial radiance is 1 $Wsr^{-1}m^{-2}$ (summarized in Table 1). Further, to check the spectral stability we have studied the electroluminescence spectral evolution with time at a constant applied current (20 mA) (supporting information S11). The consistent shape and emission peak position further demonstrate the high stability of the optimized devices in high current injection regime. To our knowledge, these figures are the best ever reported for SWIR/NIR QD based LEDs [18-21].

## 3. Conclusions

In summary, we demonstrate a new approach in improving charge balance in CQD solids by engineering the bandgap of the carrier supply-matrix QD film. By doing so we have drastically suppressed efficiency droop in CQD LEDs for the first time in the infrared and have reported EQEs as high as 8% at a radiance level of 2-10 $Wsr^{-1}m^{-2}$, the highest reported to date. Moreover, we demonstrate LEDs that present very high stability even when continuously operated in the high current injection regime. Our work therefore overcomes multiple identified challenges in the CQD LED field and demonstrates their true potential towards commercialization.

## 4. Experimental section

*Synthesis of PbS QDs:* 1.75, 1.5 and 1.32 eV excitonic peak based PbS QDs, used as the matrix materials for the devices, were synthesized using Schlenk technique following the standard recipe. Briefly, 0.446 g (2 mmol) lead oxide (PbO), 1.6 mL (4.7 mmol) oleic acid, and 18 mL 1-octadecene (ODE) were pumped overnight at 90 ºC. For 1.75 eV and 1.5 eV PbS QDs, the temperature was set to 80 °C and 1 mmol hexamethyldisilane (HMS) mixed with 5 mL ODE was immediately injected. The reaction was quenched with cold acetone after a few seconds (20 seconds for 1.75 eV PbS QDs and 50 seconds for 1.5 eV QDs) and the QDs were isolated by precipitation. The QDs were further purified by dispersion/precipitation with toluene/acetone 3 times. For 1.32 eV QDs, the temperature was set at 95°C and right after the injection, the heating was stopped (without removing the heating mantle) and the solution was allowed to cool down gradually (~1 h).

0.95 eV excitonic peak based PbS QDs were synthesized by a previously reported multi-injection method with slight modification [22]. Typically, 0.446 g (2 mmol) PbO, 3.8 mL oleic acid and 50 mL ODE were mixed together at 90 ºC under vacuum overnight. Once under Ar, the temperature of the reaction was raised to 100 ºC and a solution of 90 μL HMS in 3ml ODE was quickly injected. After 6 minutes, a second injection of 65μL of HMS in 9 mL of ODE was dropwise injected and then, the heating was stopped immediately and the solution was cooled down gradually to room temperature. The QDs were purified three times with a mixture of acetone/ethanol, redispering with toluene. For the synthesis of the 0.88 eV excitonic peak based PbS QDs, the procedure is the same but using 70 μL of HMS in both solutions. The final concentration of all the PbS QDs solutions were adjusted to 30 mg mL$^{-1}$ before the LED device preparation.

*ZnO NC synthesis:* ZnO NCs were synthesized following standard method as described in our previous report [8]. 2.95 g dihydrated zinc acetate was dissolved in 125 mL methanol under continuous stirring and the temperature of the reaction bath was set to 60 °C. In a separate vial, 1.48 g potassium hydroxide (85%) flakes were dissolved in 65 mL methanol solution. The KOH solution was then added dropwise to the zinc acetate solution for a time span of 4-5 minute while the reaction temperature was maintained around 60 °C. After 2.5 hour, the heating source was removed and the reaction bath was allowed to cool down to room temperature. The solution was then transferred to centrifuge tubes and centrifuged at 5000 rpm for 5 min. The supernatants were then discarded and remnants were washed again with methanol following the similar procedure and finally dried in nitrogen flow to get ZnO NCs. For electron injection layer, ZnO NCs were dispersed in a solution of 2% butylamine in chloroform with a concentration of 40 mgmL$^{-1}$, and to form the ternary blend with PbS QDs, ZnO NCs were dispersed in 5% butylamine in toluene solution with a concentration of 30 mgmL$^{-1}$.

*LED device preparation:* LEDs were prepared on pre-cleaned ITO coated glass as described in our previous report [8]. ZnO NCs in chloroform (40 mg ml$^{-1}$) was spun on top of the substrate with a spin speed of 4000 rpm to form the electron injection layer. The active layer was deposited on top of the ZnO layer. For blended devices, all QDs (1.75, 1.5, 1.32 and 0.95 eV excitonic peak based PbS) were prepared in separate vials with the same concentration (30 mg ml$^{-1}$) before mixing. For binary blends, emitter PbS QDs were mixed to matrix PbS QDs as 7.5% mixing ratio. For ternary blends, binary blends were further mixed with ZnO NCs. ZnI$_2$ and MPA based mixed ligand was used for the QD film formation as described in our previous report [23]. The thickness of the active layer was adjusted as ~60 nm. The hole-transporting layer of ~50 nm was formed by using the PbS QDs (similar to the matrix QDs) treated with 0.02% EDT in acetonitrile solution. The top electrode was formed by thermally evaporated Au deposition through a pre-patterned shadow mask (Nano 36 Kurt J. Lesker) at a base pressure of 10$^{-6}$ mbar. The active area of each device was 3.14 mm$^2$.

*LED performance characterization:* All the devices were fabricated and characterized in ambient air conditions. Current density–voltage (J-V) characteristics were recorded using a computer-controlled Keithley 2400 source measurement unit. EQE of the devices were calculated by detecting the radiance from the device with a calibrated Newport 918D-IR-OD3 germanium photodetector connected to Newport 1918-C power meter in parallel to the J–V measurements and further confirmed with a Newport 818 IG InGaAs photodetector. Lambertian emission was assumed. The thickness of the glass substrate was taken into account during the solid angle calculation. For stability test, the radiance of the device was constantly monitored with aforementioned photodetector with a constant applied current.

*Photoluminescence and electroluminescence measurements:* Photoluminescence (PL) and quantum yield (PLQY) measurements were performed using a Horiba Jobin Yvon iHR550 Fluorolog system coupled with a Hamamatsu RS5509-73 liquid-nitrogen cooled photomultiplier tube and a calibrated Spectralon-coated Quanta-phi integrating sphere. The excitation source for all the steady state measurement was a Vortran Stradus 637 nm continuous wave laser. The PL spectra were corrected for the system response function. The detail of the PLQY measurements was described elaborately in our previous report [8]. Electroluminescence spectra were taken using ANDOR InGaAs array CCD camera. The voltage bias to the device for electroluminescence measurement was applied with a Keithley 2400 source meter.

**Acknowledgements**

The authors acknowledge financial support from the European Research Council (ERC) under the European Union's Horizon 2020 research and innovation programme (grant agreement no. 725165), the Spanish Ministry of Economy and Competitiveness (MINECO), and the "Fondo Europeo de Desarrollo Regional" (FEDER) through grant TEC2017-88655-R. The authors also acknowledge financial support from Fundacio Privada Cellex, the program CERCA and from the Spanish Ministry of Economy and Competitiveness, through the "Severo Ochoa" Programme for Centres of Excellence in R&D (SEV-2015-0522).

**Figures:**

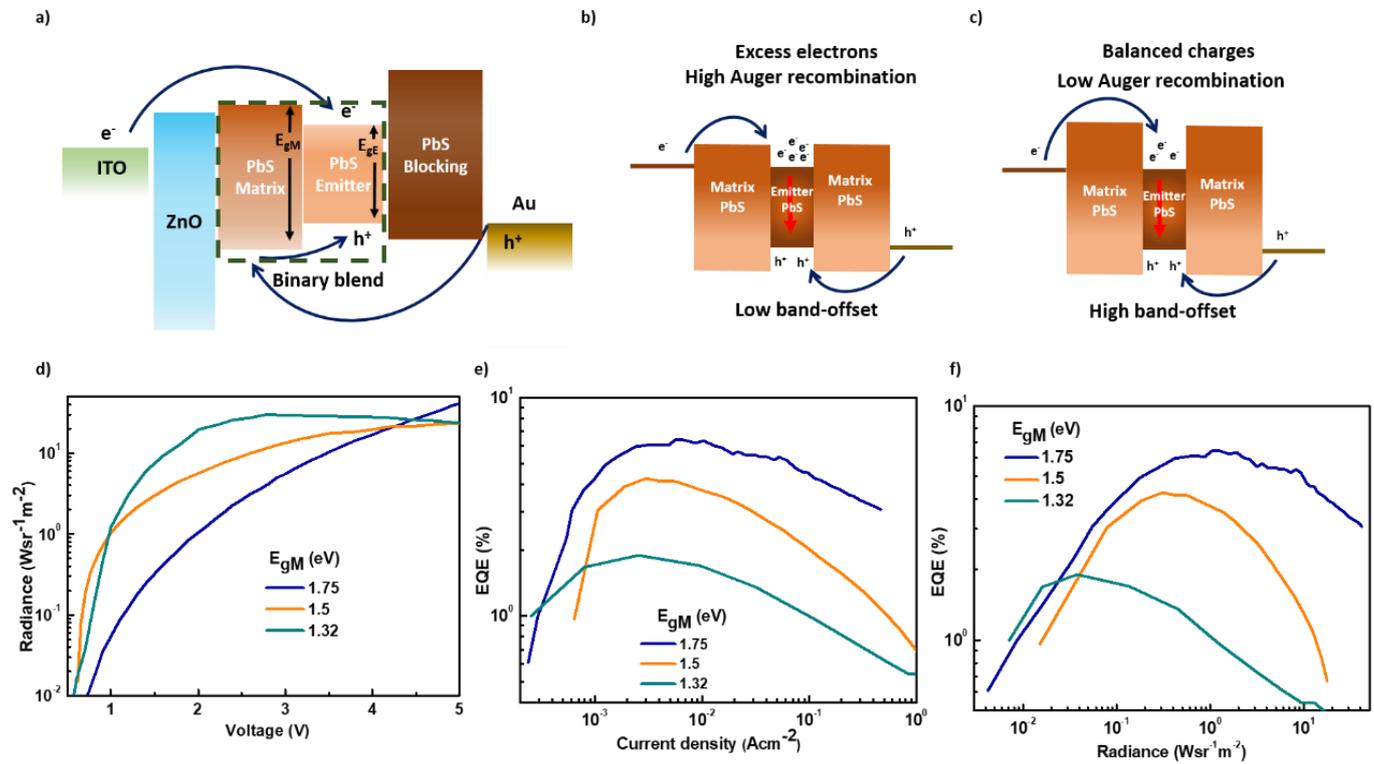

**Figure 1: The device performance as a function of matrix QD bandgap ($E_{gM}$):** (a) Schematic energy levels of binary QD blended devices. (b) & (c) The effect of energy barriers on the charge injection. The red arrow indicates the radiative recombination. The lower barrier (b) helps high number of electron injection that leads to excess electrons in the active layer. Higher energy barrier (c) regulates the electron injection and hence reduces the charge imbalance. (d) Radiance with applied voltage bias as a function of $E_{gM}$ variation in the blend. (e) EQE as a function of injected current density. Devices based on higher $E_{gM}$ show reduced EQE droop in the high current injection regime. (f) EQE as a function of radiance. Devices based on 1.75 eV $E_{gM}$ QDs show much-improved EQE in high radiance regime.

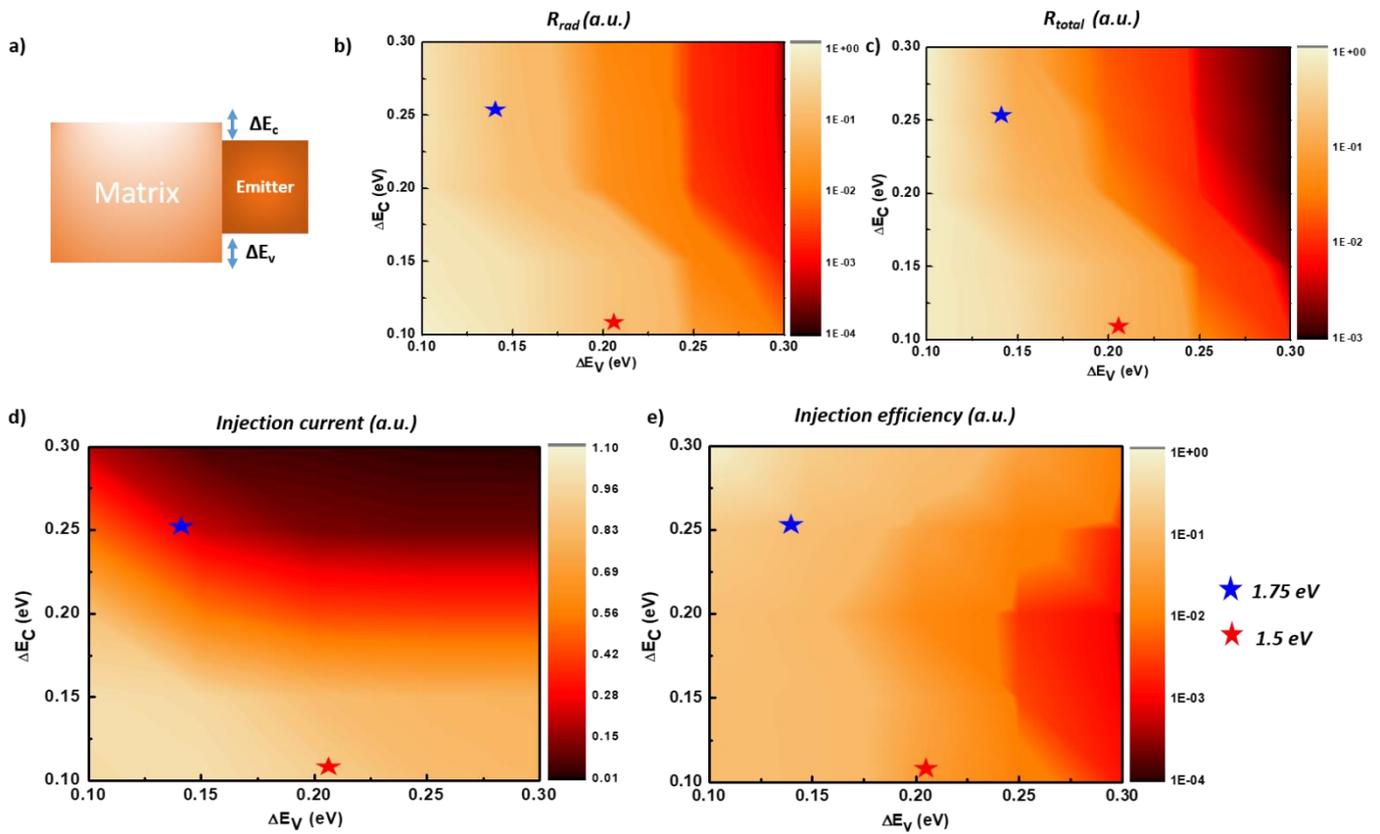

**Figure 2: SCAPS simulation to estimate the different LED performance parameters as a function of energy band offset.** (a) Schematic band levels of emitter and matrix QD heterojunction. The variation of (b) radiative recombination, (c) total recombination, (d) injection current, and (e) injection efficiency as a function of conduction ($\Delta E_C$) and valence band off-set ($\Delta E_V$). The simulation predicts a better injection efficiency for 1.75 eV $E_{gM}$ QD based devices compared to 1.5 eV $E_{gM}$ based device.

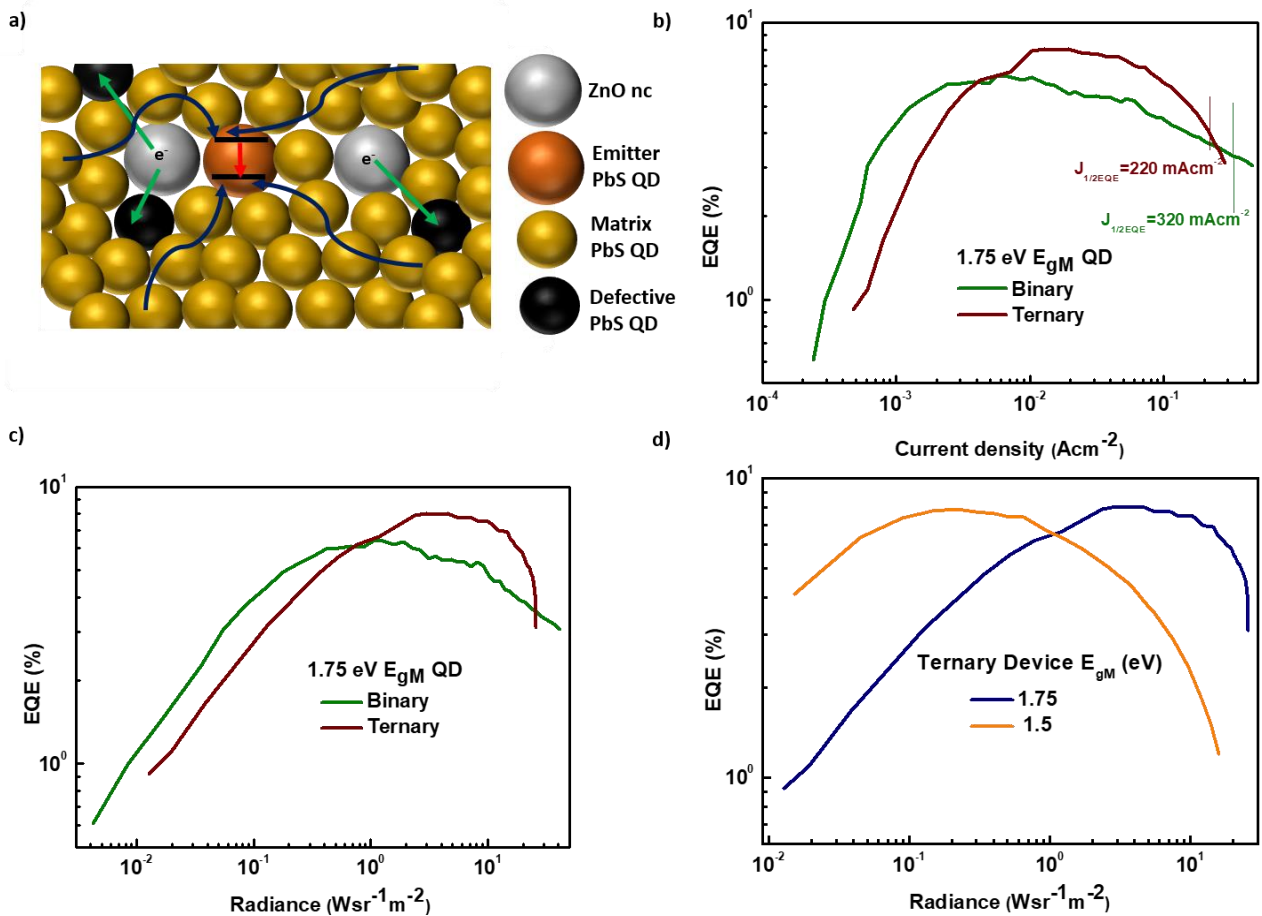

**Figure 3: Ternary device performance:** (a) Schematic of ternary blends. The ZnO NCs work as the trap passivator in the blend. It reduces the traps and trap induced non-radiative recombination. The charges are transported to the emitter QDs and radiative recombination take place at the emitter sites. Comparison of the LED EQE as a function of (b) injected charge density and (c) radiance between ternary and binary devices based on 1.75 eV $E_{gM}$ QD. Ternary devices showed better EQE in the high injection/radiance regime. (d) Comparison of the performance of ternary devices based on 1.75 eV and 1.5 eV $E_{gM}$ QDs. 1.75 eV QD based devices showed much improved performance in the high radiance region.

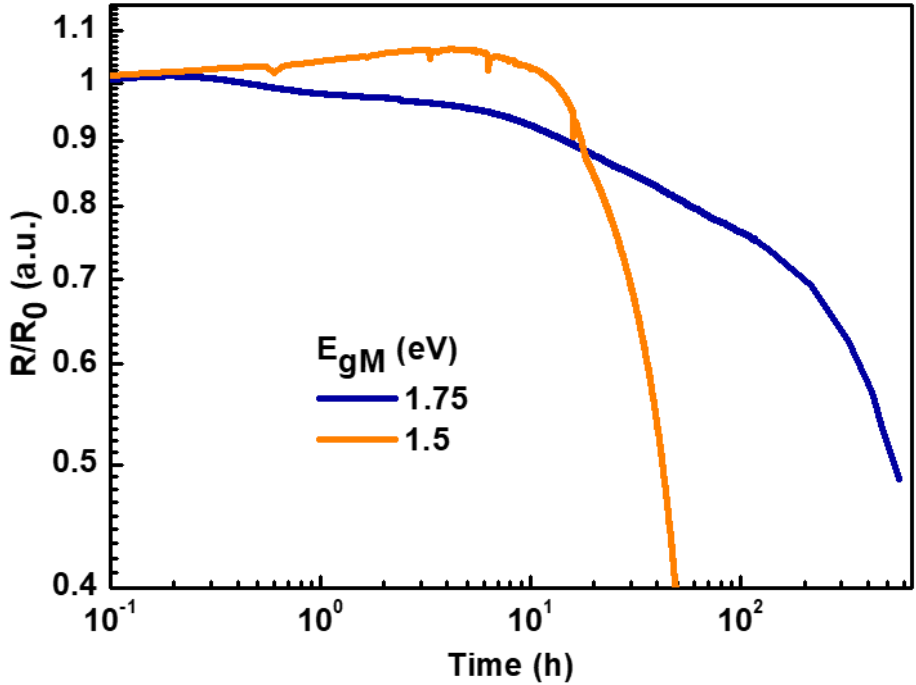

**Figure 4: LED device stability:** The comparison of the operational stability between the 1.75 eV and 1.5 eV $E_{gM}$ QD based non-encapsulated devices in air ambient conditions (60% humidity & 298 K temperature). The initial radiance was 7 $Wsr^{-1}m^{-2}$ for both the devices. Devices based on 1.75 eV $E_{gM}$ showed much higher operational stability compared to the 1.5 eV based one.

**TABLES**

| Device QD $E_{gM}$ (eV) | $t_{1/2}$ (@7 $Wsr^{-1}m^{-2}$) (expt.) (h) | $t_{1/2}$ (@1 $Wsr^{-1}m^{-2}$) (estimated) (h) | $t_{1/2}$ (@ 0.1 $Wsr^{-1}m^{-2}$) (estimated) (h) |
|---|---|---|---|
| 1.75 | 532 | 26068 | 2606800 |
| 1.5 | 42 | 2058 | 205800 |

**Table 1:** The comparison of operation device stability of the binary blend devices based on 1.75 and 1.5 eV $E_{gM}$ QD based device.